\documentclass[aps,prb,twocolumn,footinbib,superscriptaddress,floatfix]{revtex4-2}
\usepackage{amsmath}
\usepackage{graphicx}
\usepackage{indentfirst}
\usepackage{physics}
\usepackage{braket}
\usepackage{float}
\usepackage{mathtools}
\usepackage{epstopdf} 
\usepackage{footnote}
\usepackage{esint}
\usepackage{comment}
\usepackage{color}
\usepackage[T1]{fontenc}
\usepackage[caption=false,position=top]{subfig}
\usepackage{amsfonts}
\usepackage{footmisc}
\usepackage{scrextend}
\usepackage{multirow}
\usepackage[hyperfootnotes=false]{hyperref}
\usepackage[acronym]{glossaries}
\usepackage[english]{babel}
\usepackage{url}
\usepackage{bm}
\usepackage{algpseudocode}
\usepackage{tikz}
\usepackage{etoolbox}
\usepackage{enumerate}
\usepackage{soul}
\usepackage{amssymb}
\definecolor{darkblue}{rgb}{0,0,0.5}
\hypersetup{
    colorlinks=true,
    linkcolor=black,
    filecolor=blue,
    citecolor=darkblue,  
    urlcolor=black,
}

\apptocmd{\sloppy}{\hbadness 9999\relax}{}{}

\newtheorem{theorem}{Theorem}

\newtheorem{lemma}[theorem]{Lemma}

\def\be{\begin{equation}}
\def\ee{\end{equation}}
\def\ba{\begin{eqnarray}}
\def\ea{\end{eqnarray}}

\newcommand{\QZ}[1]{{{\textcolor{black}{#1}}}}
\newcommand{\JJ}[1]{{{\textcolor{black}{#1}}}}

\begin{document}

%\title{Thermalization and Many‑Body Zeno Effect in Monitored Hamiltonian Dynamics}
%\title{Determinization,  Derandomization, Confinement}
\title{\QZ{Many-Body Anti-Zeno Thermalization and Zeno Determinism \\
in Monitored Hamiltonian Dynamics}}
\author{Jia-Jin Feng}
\email{jiajinfe@usc.edu}
\affiliation{
Ming Hsieh Department of Electrical and Computer Engineering, University of Southern California, Los
Angeles, California 90089, USA
}

\author{Quntao Zhuang}
\email{qzhuang@usc.edu}
\affiliation{
Ming Hsieh Department of Electrical and Computer Engineering, University of Southern California, Los
Angeles, California 90089, USA
}
\affiliation{ Department of Physics and Astronomy, University of Southern California, Los
Angeles, California 90089, USA
}

\begin{abstract}
Random quantum states are essential for quantum information science, with applications ranging from quantum computing to cryptography. Prior approaches for generating these states often rely on using a large bath to thermalize a smaller system, with a subsequent measurement on the bath used to post-select a random state. 
To reduce the required size of the bath, we propose a resource-efficient scheme using holographic deep thermalization driven by Hamiltonian evolution, combined with mid-circuit measurements. \JJ{This scheme relies on dynamical circuits, enabling a trade-off between spatial and temporal resources and allowing the generation of genuinely random states with only a constant-size bath.} We quantify the randomness using the frame potential and derive its asymptotic behavior, which shows good agreement with our numerical simulations \QZ{and experimental results on IBM quantum devices}. \JJ{For a fixed total evolution time, increasing the number of mid-circuit measurements initially produces an exponential decrease in the frame potential---a quantum anti-Zeno behavior arising from holographic deep thermalization.} Past a critical number of mid‑circuit measurements, the frame potential rises again, signaling the onset of the quantum Zeno effect.
\end{abstract}

\maketitle

\section{Introduction}

\QZ{Quantum systems equilibrate and thermalize~\cite{reimann2008foundation,rigol2008thermalization,short2011equilibration,rigol2012alternatives,short2012quantum} through chaotic and entangling unitary interaction with a large bath~\cite{popescu2006entanglement}, causing its reduced state to appear thermal~\cite{Donovan2025,Mori2023}.
More recently, the notion of deep thermalization~\cite{ho2022exact,cotler2023emergent} further considers the state ensemble generated by quantum measurement on the bath, given that the system-bath state is pure. }Conditioned on the measurement result, the system is in a random pure state~\cite{Giulia2025}; collecting conditional states under the different measurement outcomes, one therefore obtains an ensemble of pure states for the system. Refs.~\cite{ho2022exact,cotler2023emergent} show that a measurement-induced ensemble becomes Haar random when the unitary dynamics of the system and the bath are sufficiently intricate. 

While a large bath compared to the system is essential for deep thermalization~\cite{ho2022exact,cotler2023emergent} and quantum thermalization~\cite{reimann2008foundation,rigol2008thermalization,short2011equilibration,rigol2012alternatives,short2012quantum}, a small bath with certain dynamical control also plays an important role in statistical physics~\cite{metcalf2020engineered}, \QZ{as exemplified by the famous “Maxwell’s demon”~\cite{ciccarello2022quantum}.}
\QZ{
Via repeated interaction, measurement and reset, holographic deep thermalization(HDT)~\cite{zhang2025holographic} further allows a small bath to enable Haar random pure states on the system.} In the initial proposal~\cite{zhang2025holographic}, however, each unitary interaction between the system and bath is assumed to be a fixed unitary sampled from Haar ensemble. 
While effective for circuit-based platforms, this assumption leaves open the question of how HDT can be faithfully realized using continuous Hamiltonian dynamics. In particular, implementing Haar-typical unitaries typically requires long evolution times, which may present constraints for experimental realizations.

In this work, we extend the study of holographic deep thermalization to Hamiltonian dynamics and uncover a subtle interplay between anti-Zeno thermalization and the quantum Zeno effect. 
By employing an interacting Hamiltonian and performing intermediate measurements on the bath, we aim to generate Haar-random states.
\QZ{Given a fixed evolution time,} our findings show that increasing the number of intermediate measurements \QZ{initially enhances the randomness of the resulting state ensemble, in analogy to the anti-Zeno effect(AZE)~\cite{kofman2000,Facchi2001,Fischer2001,Chaudhry2014,Leppenen2022,Gumber2023,Zhang2025}.} However, beyond a certain threshold, excessive measurements trigger the \QZ{quantum Zeno effect(ZE)~\cite{misra1977zeno,FACCHI200012,facchi2022quantum,mccusker2013experimental,schafer2014experimental,signoles2014confined,slichter2016quantum,Becker2021,zeni2025,Li2018,Biella2021}}, inhibiting the bath's evolution and diminishing the ensemble's randomness. We provide both analytical and numerical insights to determine the optimal number of measurements that maximizes Haar-randomness for a fixed evolution duration. \QZ{Our results generalize AZE and ZE to partial mid-circuit measurement and connect them to deep thermalization. }

\begin{figure}[b]
\begin{center}
\includegraphics[clip = true, width =\columnwidth]{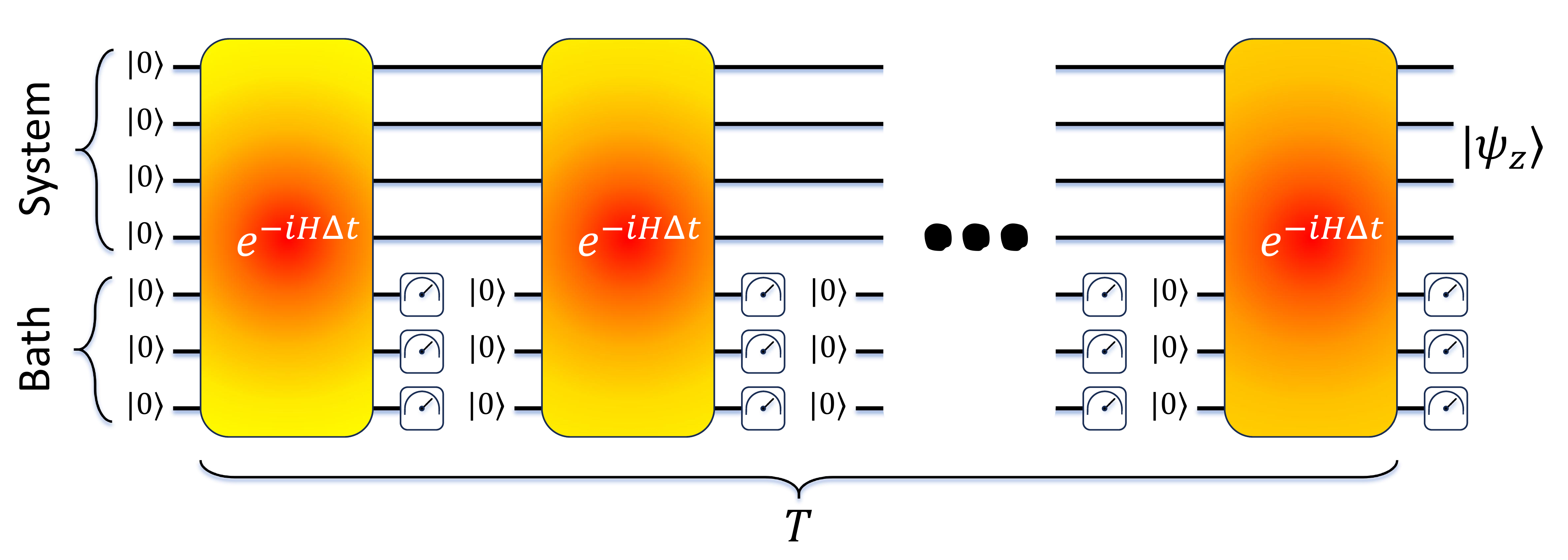}
\caption{\label{fig:Setup} Schematic of the quantum circuit implementing holographic deep thermalization via Hamiltonian dynamics with multiple mid-circuit measurements.}
\end{center}
\end{figure}

\section{Overview}
We utilize a Hamiltonian that satisfies the eigenstate thermalization hypothesis (ETH) to drive the system toward Haar randomness. The full system comprises a system register of $n_{\rm s}$ qubits and a bath of $n_{\rm b}$ ancillary qubits, for a total of $n_{\rm all}=n_{\rm s}+n_{\rm b}$ qubits. \QZ{As shown in Fig.~\ref{fig:Setup}}, the total evolution time is fixed at $T$, during which we insert $n$ intermediate measurements on the bath qubits. 
The bath is measured and then reset after each segment.
\QZ{We quantify the randomness of the measurement-generated pure-state ensemble using the frame potential.}

\begin{figure}[tb]
\begin{center}
\includegraphics[clip = true, width =\columnwidth]{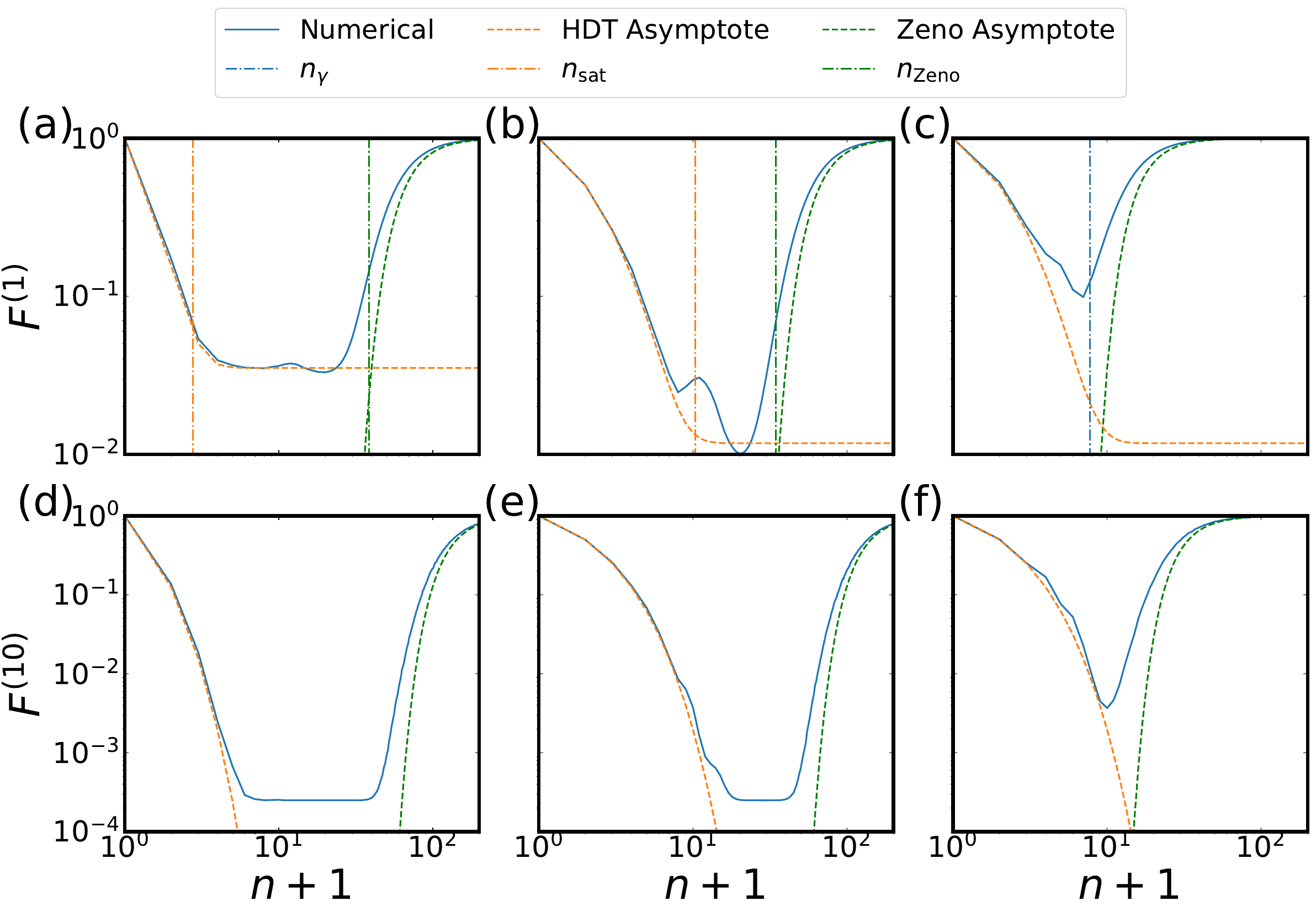}
\caption{\label{fig:Fn} Frame potential as a function of the number of measurements. The blue lines represent numerical results, while the orange and green lines correspond to analytical results for the asymptotic region of HDT in Eq.~\eqref{eq:HDT1} and the ZE in Eq.~\eqref{eq:zeno}, respectively. Other parameters are: (a),(d) $n_{\rm s}=5$, $n_{\rm b}=3$; (b),(c),(e),(f) $n_{\rm s}=7$, $n_{\rm b}=1$. Total evolution time is $T = 15/J_{zz}$ for (a),(b),(d),(e) and $T = 5/J_{zz}$ for (c),(f). The orange, green, and blue dashed lines indicate the positions of $n_{\rm sat}$, $n_{\rm Zeno}$, and $n_{\gamma}$, respectively, with $r=0.1$. Other parameters are $J_z=0.9J_{zz}$ and $J_x=1.4J_{zz}$. }
\end{center}
\end{figure}

For the Hamiltonian evolution process shown in Fig.~\ref{fig:Setup},  the frame potential initially decreases and then increases depending on the number of measurements $n$, as illustrated by the blue curves in Fig.~\ref{fig:Fn}.
The initial decrease in the small-$n$ \QZ{AZE} regime is well captured by the HDT theory \QZ{(orange curves)}.
In contrast, the increase in the large-$n$ regime can be explained by the quantum ZE \QZ{(green curves)}.
These theoretical descriptions hold across different bath sizes, as demonstrated by the comparison between cases with a small number of bath qubits in Figs.~\ref{fig:Fn}(a)(d) and those with a larger number of bath qubits in Figs.~\ref{fig:Fn}(b)(c).

When the total evolution time $T$ is large, the frame‑potential decrease saturates once $n$ approaches $n_{\rm sat}\sim O\left(\left(n_{\rm s}-{\rm log}_2 r\right)/n_{\rm b}\right)$, as indicated by the orange dashed lines in Fig.~\ref{fig:Fn}(a)(b), where $r$ is a Hamiltonian dependent parameter. Beyond this point, the frame potential first saturates to a plateau, with fluctuations induced by dynamical revival from finite-system-size effect~(Appendix \ref{app:revival}). 
After the plateau, the frame potential begins to increase due to the onset of the quantum ZE, occurring around $n_{\rm Zeno}\sim O\left(\sqrt[\alpha]{T^{\alpha+1}/n_{\rm s}}\right)$, as indicated by the green dashed lines in Fig.~\ref{fig:Fn}(a)(b), where $\alpha$ is another Hamiltonian-dependent parameter.

For small $T$ with $n_{\rm sat}>n_{\rm Zeno}$ (Fig.~\ref{fig:Fn}(c)(f)), HDT and the ZE compete, and the minimum frame potential is reached near $n_\gamma \sim O\left( T/ \sqrt[\alpha+1]{n_{\rm b}}\right)$ which reflects the optimal balance between thermalization and measurement-induced suppression.

\section{Set-up}In HDT, both the system and bath are initialized in a trivial state that is easy to prepare; here, we choose the all-zero state, denoted as $|\psi_0\rangle\otimes|\phi_0\rangle$. The system and bath are then entangled through evolution \QZ{$U=e^{-iH\Delta t}$} under a time-independent Hamiltonian $H$ \QZ{for a time interval $\Delta t$}. For  brevity, we set $\hbar=1$. After each evolution step, the bath is measured in the computational basis  ${|\phi_m\rangle}$ and reset to the trivial state $|\phi_0\rangle$. The cycle of Hamiltonian evolution, measurement, and reset is repeated $n$ times, with the total evolution time fixed at $T$. At the end of the process, the system collapses into one of the possible states $|\psi_z\rangle$ with probability $p_z$, where $z$ indexes all possible measurement trajectories. These outcomes define an ensemble $\mathcal{E}=\{p_z, |\psi_z\rangle \}$.

\JJ{The randomness of the ensemble $\mathcal{E}$ is characterized by the $K$-th frame potential~\cite{ippoliti2023dynamical}}
\begin{eqnarray}
F^{(K)}=\sum_{z,z'}p_z  p_{z'}\left| \langle \psi_z | \psi_{z'} \rangle \right|^{2K} ~,
\label{eq:fp}
\end{eqnarray}
\QZ{which measures the average overlap between different output states in the ensemble. A smaller frame potential indicates greater randomness.}
\JJ{For comparison, the Haar ensemble provides the benchmark for maximal randomness. In this ideal case, the $K$-th frame potential takes the closed form}
\begin{eqnarray}
    F_{\rm Haar}^{(K)}=\frac{(N_{\rm s}-1)!K!}{(N_{\rm s}+K-1)!} ~,
\end{eqnarray}
where $N_{\rm s}=2^{n_{\rm s}}$ is the size of system Hilbert space.

In prior work on HDT, the transverse and longitudinal-field Ising model has been used to approximate Haar-random dynamics \cite{cotler2023emergent}. In our approach, we adopt the same spin-chain Hamiltonian with open boundary condition to implement HDT, which is given by
\begin{eqnarray}
H_{\rm Ising}=J_{x}\sum_{j=0}^{n_{\rm all}-1}\sigma_j^{x}+J_{z}\sum_{j=0}^{n_{\rm all}-1}\sigma_j^{z}+J_{zz}\sum_{j=0}^{n_{\rm all}-2}\sigma^{z}_j\sigma_{j+1}^{z} ~, \nonumber\\
\label{eq:Ising}
\end{eqnarray}
where $J_{x}$ denotes the strength of the transverse field, $J_{z}$ the longitudinal field, and $J_{zz}$ the nearest-neighbor interaction strength.
%The results can be seen in Fig.~\ref{fig:Fn}

For the more general cases, we introduce strong perturbation terms to the Hamiltonian in order to disrupt any potential underlying structure. Specifically, we consider the following two types of perturbations
\begin{eqnarray}
H_{YY}&=&H_{\rm Ising}+J_{yy}\sum_{j=0}^{n_{\rm all}-2}\sigma^{y}_j\sigma_{j+1}^{y} ~, \\
H_{XXX}&=&H_{\rm Ising}+J_{xxx}\sum_{j=1}^{n_{\rm all}-2}\sigma_{j-1}^{x}\sigma^{x}_j\sigma_{j+1}^{x} ~.
\end{eqnarray}

Although the Hamiltonian may possess symmetries, such as inversion or translational symmetry, which constrain the spreading of the wave function across the full Hilbert space, the measurement and subsequent reset to the bath partially break these symmetries. This symmetry breaking, in turn, enhances both randomness and thermalization.

\begin{figure}[tb]
\begin{center}
\includegraphics[clip = true, width =\columnwidth]{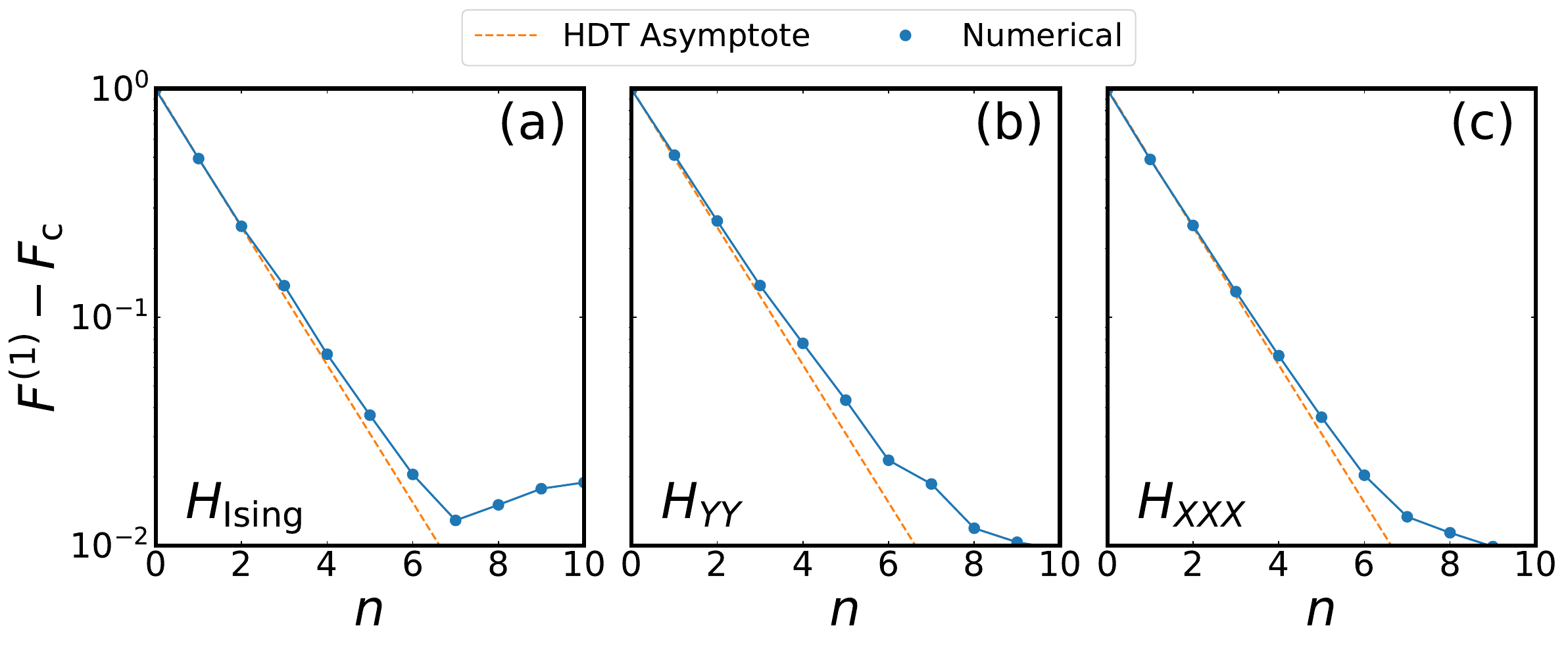}
\caption{\label{fig:HDTint} The convergence dynamics of frame potential as a function of the number of
measurements for (a) $H_{\rm Ising}$, (b) $H_{YY}$, and (c) $H_{XXX}$. 
%y axis is the frame potential difference with the converged value in Eq.~\eqref{Fc}.
Blue lines represent numerical results, while orange lines correspond to analytical predictions for HDT. Other parameters are $n_{\rm s}=7$, $n_{\rm b}=1$, $J_z=0.9J_{zz}$, $J_x=1.3J_{zz}$, and $T=15/J_{zz}$.}
\end{center}
\end{figure}

\section{Holographic deep thermalization}
In the context of HDT, the insertion of multiple mid-circuit measurements significantly accelerates the thermalization process, potentially leading to exponential speedup in approaching Haar randomness.
If the unitaries are randomly sampled from the Haar measure, the expected first-order frame potential in HDT is given by \cite{zhang2025holographic}
\begin{eqnarray}
    F^{(1)}=q_{\rm 1}+\frac{(N_{\rm s}-1)(N_{\rm s}N_{\rm b}-1)}{N_{\rm s}^2 N_{\rm b}+1} \left( \frac{(N_{\rm s}^2-1)N_{\rm b}}{N_{\rm s}^2 N_{\rm b}^2-1} \right)^n ,
\label{eq:HDT1}
\end{eqnarray}
where the corrected saturation value is
$
q_{\rm 1}\equiv \frac{N_{\rm s}^2(N_{\rm b}+1)}{N_{\rm s}^2 N_{\rm b}+1} F_{\rm Haar}^{(1)}.
$
For the higher order frame potential ($K\geqslant2$), a lower bound can be obtained as
\begin{eqnarray}
    F^{(K)}=(1-q_K)\left( \frac{N_{\rm s}^2N_{\rm b}}{N_{\rm s}^2 N_{\rm b}^2-1} \right)^n+q_K ~,
\end{eqnarray}
where the corrected saturation value is
$
    q_K=\left(1+{(2^K-1)}/{N_{\rm b}}\right)F_{\rm Haar}^{(K)}.
$

Although the \QZ{spatially-local} Hamiltonian evolution differs from random Haar-sampled unitaries, \QZ{the resulting frame potential dynamics} agrees well with the Haar unitary results when the number of measurements $n$ is small. As illustrated by the blue dotted lines in Fig.~\ref{fig:HDTint}, the frame potential exhibits a rapid decay in this regime. This behavior is well captured by the HDT theory described in Eq. (\ref{eq:HDT1}), which predicts an exponential decrease, as shown by the orange lines in Fig.~\ref{fig:HDTint}. Notably, this agreement holds across different Hamiltonians, as demonstrated in \QZ{the three subplots} of Fig.~\ref{fig:HDTint}.

Phenomenologically, when the frame potential approaches the same order as its minimal value, $F^{(1)}=(r+1)F_{\rm c}$ with small constant $r$, saturation starts to kick in. The number of measurements required for saturation is
\begin{eqnarray}
    n_{\rm sat}\approx\frac{n_{\rm s}-{\rm log}_2 r}{n_{\rm b}} ~.
\end{eqnarray}

After the saturation of the frame potential, small fluctuations remain. These are primarily due to finite-size effects and the limited evolution time in the Hamiltonian dynamics. A detailed explanation can be found in Appendix \ref{app:revival}.

\begin{figure}[tb]
\begin{center}
\includegraphics[clip = true, width =\columnwidth]{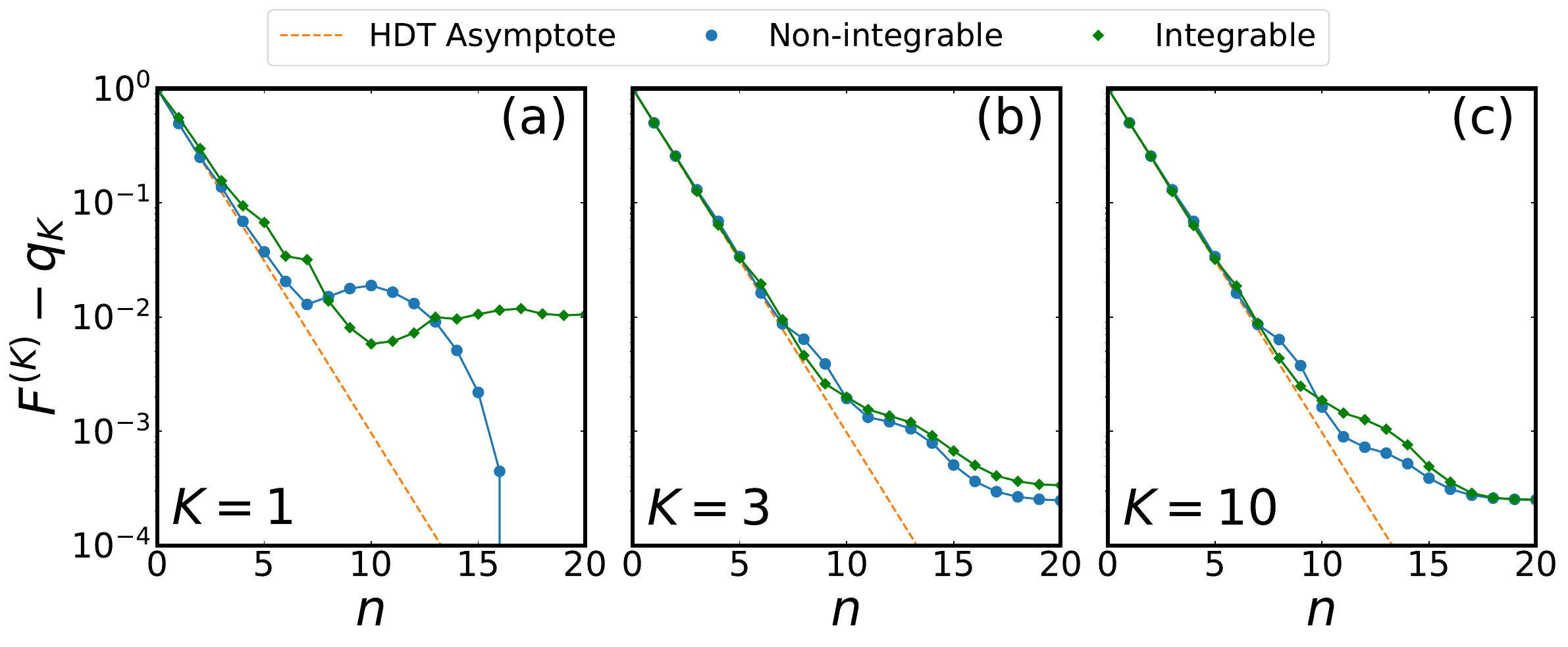}
\caption{\label{fig:HDTK} Frame potential for $K$-designs corresponding to (a) $K=1$, (b) $K=3$, and (c) $K=10$. The black solid line represents the analytical approach. Blue dots indicate results from the non-integrable Hamiltonian ($J_z=0.9J_{zz}$), while green dots correspond to the integrable Hamiltonian ($J_z = 0$). Other parameters are $n_{\rm s}=7$, $n_{\rm b}=1$, $J_x=1.3J_{zz}$, and $T=15/J_{zz}$~. }
\end{center}
\end{figure}

The dynamics of integrable and non-integrable systems are fundamentally different, and this distinction is also reflected in the behavior of the low-order frame potential. By tuning the Hamiltonian in Eq. \eqref{eq:Ising}, we can control its integrability. Specifically, setting $J_z=0$ renders the system integrable, whereas a nonzero $J_z$ leads to non-integrable dynamics.

In the regime of a small number of measurements 
$n$, the non-integrable system more closely follows the predictions of HDT, while the integrable system deviates more rapidly from the HDT curve, as shown in Fig.~\ref{fig:HDTK}(a). As $n$ increases, both systems exhibit oscillatory behavior. However, as the order of the frame potential increases, the distinction between integrable and non-integrable systems becomes less pronounced, as illustrated in Fig.~\ref{fig:HDTK} (b) and (c).

For higher-order frame potentials, both systems saturate at similar values, and the oscillations gradually vanish as the design increases. This suggests that at high design, the distinction between integrable and non-integrable local Hamiltonians becomes negligible.

We find that both integrable and non-integrable local Hamiltonians remain far from the minimum value predicted by HDT theory, as shown in Fig.~\ref{fig:HDTK}(c). This deviation is likely due to the inherent locality of the Hamiltonian. Although a chaotic Hamiltonian may exhibit ergodic behavior in local observables, it does not fully explore the entire Hilbert space. One fundamental limitation is energy conservation, which restricts dynamics to a finite energy shell. Moreover, due to locality constraints, the evolution cannot uniformly cover the Hilbert space in the same way as a Haar random unitary. Furthermore, the Hamiltonian contains only a limited number of independent parameters, meaning that the matrix elements of the evolution operator $U=e^{-iH\Delta t}$  are not fully independent. In contrast, a Haar-random unitary has a number of independent parameters equal to the dimension of the Hilbert space minus one. Consequently, the dynamics generated by a local Hamiltonian cannot reach the level of randomness characteristic of a typical Haar-random unitary.
These observations highlight the intrinsic limitations of using local Hamiltonians and restricted resources to generate high-order quantum randomness.
 
\begin{figure}[t]
\begin{center}
\includegraphics[clip = true, width =\columnwidth]{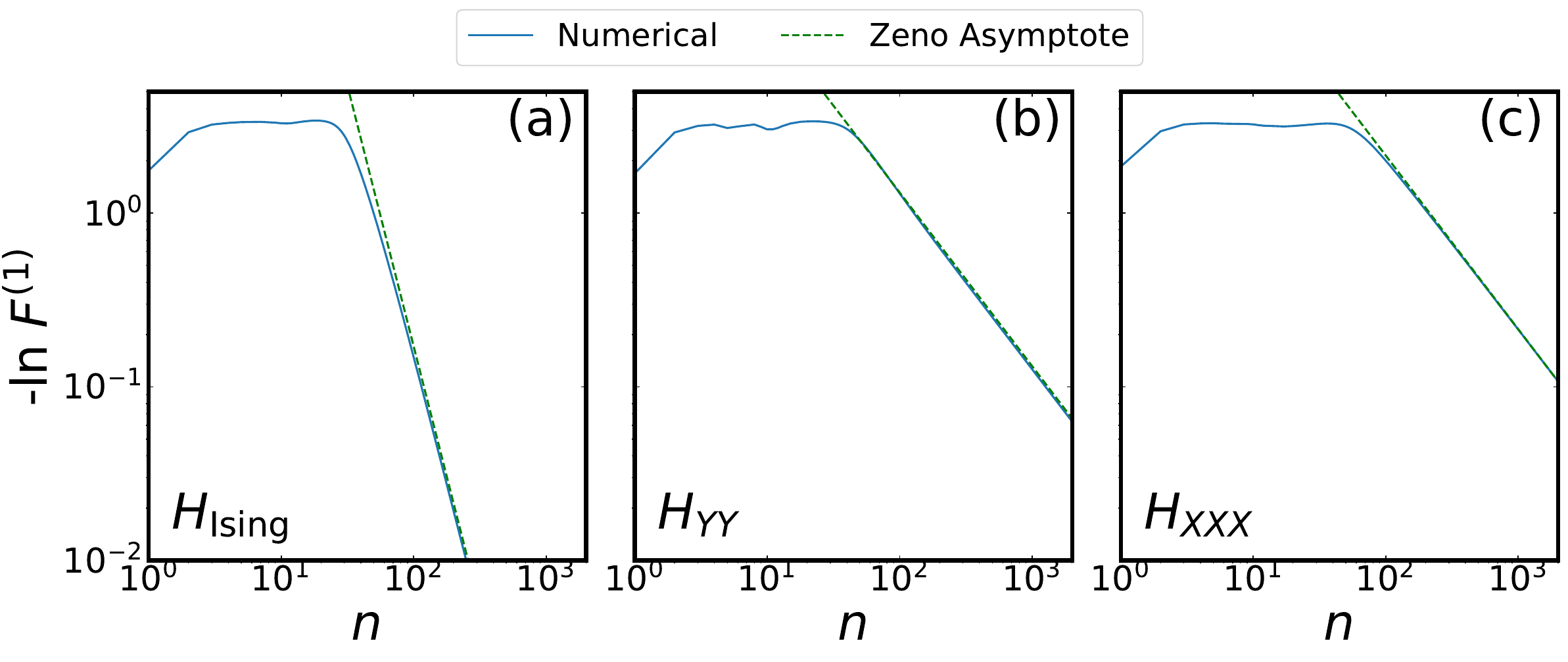}
\caption{\label{fig:zeno} Frame potential as a function of the number of measurements for (a) $H_{\rm Ising}$, (b) $H_{YY}$, and (c) $H_{XXX}$. Blue lines represent numerical results, while green lines correspond to analytical predictions for ZE with (a) $\alpha=3$, and (b)(c) $\alpha=1$. The total evolution time is  $T=15/J_{zz}$. The system consists of 5 qubits, with 3 bath qubit. Other parameters are $J_z=0.9J_{zz}$ and $J_x=1.3J_{zz}$.}
\end{center}
\end{figure}

\section{Zeno effect}
The coupling of the system to the bath can give rise to the ZE \cite{Maimbourg2021,Heinrich2020}.
When measurements on the bath are performed excessively often, the quantum ZE can effectively freeze the bath, suppressing its intrinsic randomness.
For a finite evolution time $T$, if the number of measurements $n$ is too large, the bath will remain almost fixed in its reset state $|\phi_0\rangle$. Consequently, the probability of obtaining any measurement outcome other than the reset state approaches zero. The system therefore remains in a pure state, with the frame potential approaching one.

The short-time dynamics can be analytically calculated in the limit of $\Delta t\rightarrow 0$, which allows us to investigate the asymptotic behavior of the frame potential.

\begin{lemma}
For a fixed total evolution time $T$, in the limit of $n \gg 1$, we obtain the following asymptotic lower bounds for the frame potential
 
\begin{eqnarray}
    F^{(K)} \gtrsim e^{-O\left(\frac{T^{\alpha+1}}{n^\alpha}  \right)} ~, 
\label{eq:zeno}
\end{eqnarray}
\QZ{where $\alpha$ is a constant.}
\end{lemma}
\QZ{In general, the constant $\alpha=1$; while a tighter bound with $\alpha=3$ arises in the presence of a special symmetry. The corresponding deviations are provided in the Appendix \ref{app:Zeno}. }
When $n$ is large, the frame potential approaches 1. To extract the exponential decay behavior, we plot the data on a logarithmic scale. The asymptotic behavior of the frame potential for different Hamiltonians is shown in Fig.~\ref{fig:zeno}. The numerical results are represented by the blue curves, while the exponential decay rate given in Eq. \eqref{eq:zeno} is indicated by the green lines. As $n$ increases, the numerical results converge more closely to the predicted exponential behavior.

\JJ{Combining Eq. (\ref{eq:HDT1}) with Eq. (\ref{eq:zeno}), we can estimate the measurement threshold at which the ZE becomes observable. This threshold is approximately}
\begin{eqnarray}
    n_{\rm Zeno}&\approx&\sqrt[\alpha]{\frac{c_HT^{\alpha+1}}{n_{\rm s}\ln 2}},
    \label{n_Zeno}
\end{eqnarray}
\JJ{where $c_H$ is a characteristic constant of the Hamiltonian associated with the wave function's diffusion and scrambling rates. It quantifies the ability of the system's back-action on the bath. The derivation of $c_H$ is provided in the Appendix \ref{app:Zeno}.}
%Appendix \ref{app:Zeno}.
The number of measurements should not exceed this value to preserve states with a good level of randomness.

Saturation of the frame potential requires that it occur before the onset of the ZE, i.e., $n_{\rm sat}<n_{\rm Zeno}$. In other words, the total evolution time must satisfy
\begin{eqnarray}
    T> \sqrt[\alpha+1]{\frac{(n_{\rm s}-{\rm log}_2 r)^\alpha n_{\rm s} \ln 2}{n_{\rm b}^\alpha c_H}} ~.
\end{eqnarray}
To achieve a Haar-random ensemble, the total Hamiltonian evolution time should be greater than this threshold.

When the total evolution time $T$ is sufficiently short such that $n_{\rm sat}>n_{\rm Zeno}$, there is a competition between HDT and ZE. In this regime, the frame potential does not reach its saturation value and remains significantly above it. Combining Eq. (\ref{eq:HDT1}) and Eq. (\ref{eq:zeno}), the minimum frame potential is attained at around
\begin{eqnarray}
    n_\gamma \approx T \sqrt[\alpha+1]{\frac{c_H}{n_{\rm b}\ln 2}}.
\end{eqnarray}
This corresponds to the recommended number of measurements under a limited total evolution time, constrained by experimental factors like coherence time.

\begin{figure}[tb]
\begin{center}
\includegraphics[clip = true, width =\columnwidth]{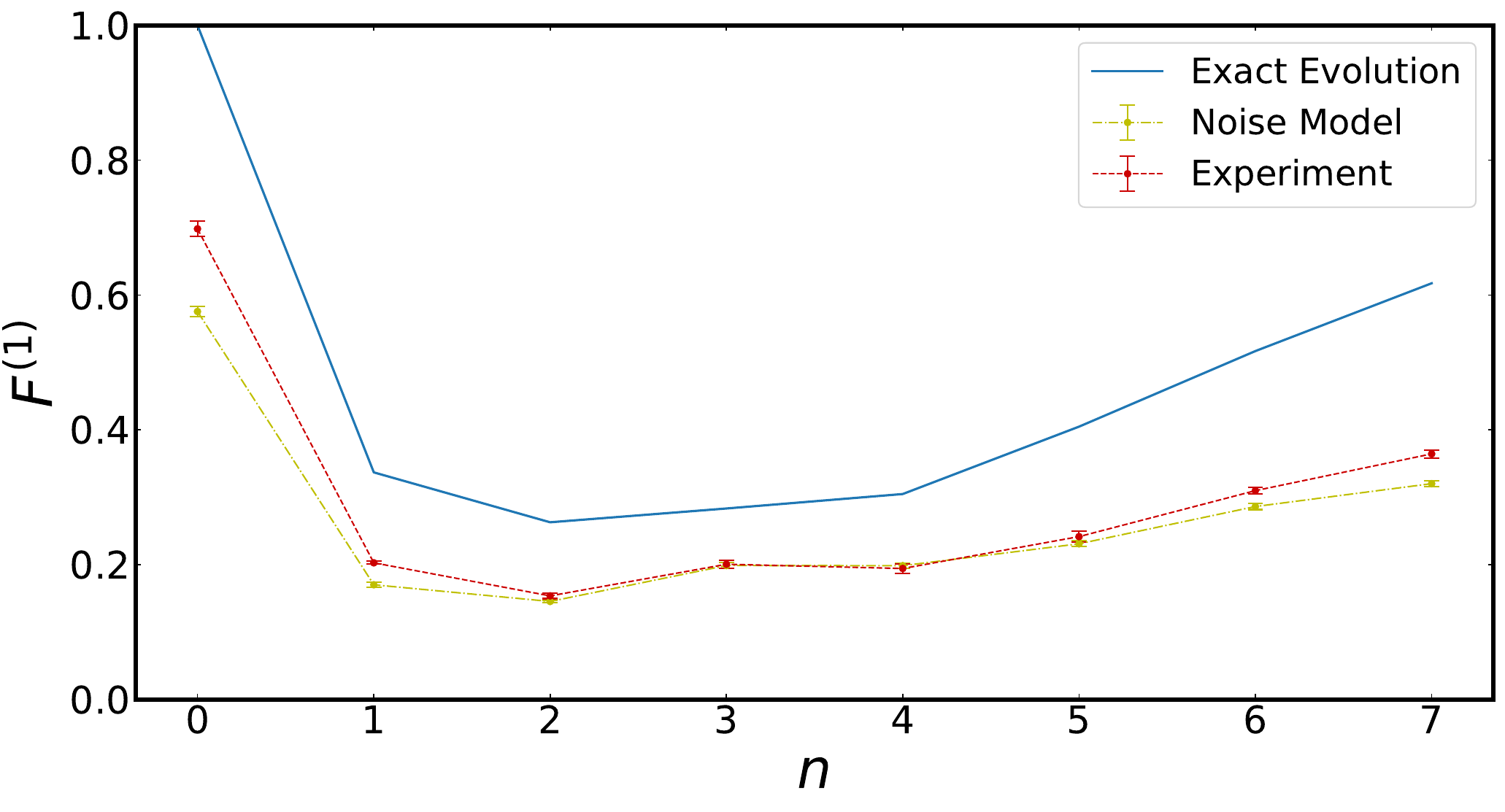}
\caption{\label{fig:Exp} Experimental results on frame potential versus the number of mid-circuit measurements. The dynamics are generated by  $H_{\rm Ising}$ and executed experimentally on the QPU of "ibm\_torino". 
%The noisy simulations use the built-in noise model, with parameters extracted directly from the same device. 
Error bars denote the standard deviation across independent data ensembles. The total evolution time is $T=3/J_{zz}$. The system contains 3 system qubits coupled to 3 bath qubits. Other parameters are $J_z=0.9J_{zz}$ and $J_x=1.4J_{zz}$.}
\end{center}
\end{figure}

\section{ Experimental results} 
In experiments, mid-circuit measurement is crucial for implementing quantum error correction~\cite{google2025quantum} and realizing dynamical circuits~\cite{Baumer2024,Baumer20242}. Such capabilities enable scaling to larger system sizes and support more efficient quantum computation ~\cite{cao2025}.

We performed a proof-of-principle experiment on a small-scale system using the IBM Quantum Platform. Hamiltonian evolution is realized through a Suzuki–Trotter decomposition. The first-order frame potential $F^{(1)}$ can be extracted via full state tomography on the system qubits for the system size in our study~(Appendix \ref{app:Experiment}). As shown in Fig.~\ref{fig:Exp}, the experimental data (red dots), exhibit good agreement with the theoretical prediction (blue solid line), with a finite reduction attributable to experimental noise, consistent with the corresponding noisy simulation (yellow dash-dotted line).

\section{Conclusion}We propose a protocol based on multiple Hamiltonian evolutions, mid-circuit measurements, and resets to realize holographic deep thermalization. This approach reduces the number of required bath (ancilla) qubits and relaxes the need for large-scale qubit entanglement in experiments aimed at generating quantum randomness.

\begin{acknowledgments}
The authors acknowledge discussions with Bingzhi Zhang. The project is supported by Office of Naval Research Grant No. N00014-23-1-2296 and DARPA HR0011-24-9-0362. QZ also acknowledges support from NSF (OMA-2326746, 2350153, CCF-2240641), AFOSR MURI FA9550-24-1-0349, DARPA HR0011-24-9-0362, Halliburton Company, an unrestricted gift from Google. 
\end{acknowledgments}
\appendix

\section{Revival dynamics}
\label{app:revival}
As the frame potential approaches its saturation value, small fluctuations appear, primarily arising from revivals in the Hamiltonian evolution of the finite-size system. In this regime, the energy spectrum becomes discrete with only a small number of energy levels, and the smoothing effect of thermalization diminishes. 

\begin{figure}[b]
\begin{center}
\includegraphics[clip = true, width =\columnwidth]{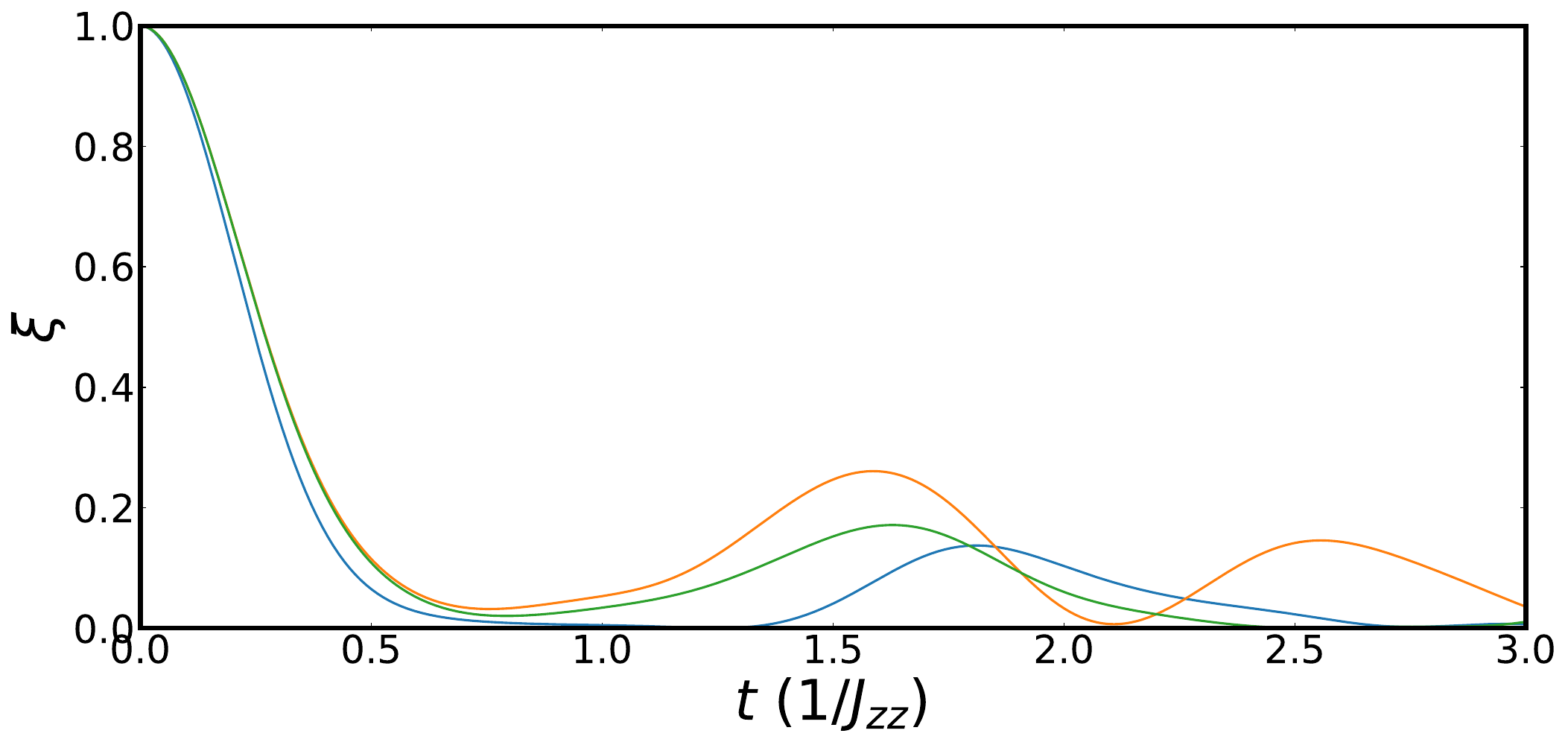}
\caption{\label{fig:Revival} Fidelity of representative states during the time-dependent revival dynamics governed by $H_{\rm Ising}$.  The system comprises 7 qubits and 1 bath qubit. }
\end{center}
\end{figure}

To quantify these revivals, we use the fidelity between the evolved state $\psi_t$ and the initial state $\psi_0$, defined as
\begin{eqnarray}
    \xi=\left| \langle \psi_0|\psi_t \rangle  \right|^2.
\end{eqnarray}
In Fig. \ref{fig:Revival}, we show the fidelity of different representative states as a function of time. The first revival occurs at approximately 1.6 $\mu$s, where the fidelity no longer decreases but instead rises to a small peak. In Fig. 2(b) of the manuscript, the frame potential similarly stops decreasing and exhibits a small peak near the time interval $\Delta t = T/n = 1.5/J_{\rm zz} $. These two values are of the same order of magnitude, indicating that the fluctuation and the atypical increase of the frame potential originate from the revival in the Hamiltonian evolution.

\section{Zeno effect}
\label{app:Zeno}

In the following, we provide the derivation of Eq.~(\ref{eq:zeno}) from the main text. We consider Hamiltonian evolution in a composite system where a subset of qubits is designated as a bath and subject to repeated measurement and reset. In general, the Hamiltonian can be decomposed into three parts
\begin{eqnarray} 
H&=&H_{{\rm s}}+H_{{\rm b}}+H_{{\rm c}} \nonumber\\
&=&H_{{\rm s}0}+H_{{\rm b}}+H_{{\rm c}0} ~,
\end{eqnarray}
where $H_{{\rm s}},H_{{\rm b}}$ and $H_{{\rm c}}$ correspond to the system, bath and interaction Hamiltonian. In the second step, we introduced a modified system and interaction Hamiltonian
\begin{eqnarray}
    H_{{\rm s}0}=H_{{\rm s}}+\langle \phi_0 | H_{\rm c} | \phi_0  \rangle,
\end{eqnarray}
and
\begin{eqnarray}
    H_{{\rm c}0}= H_{\rm c} -\langle \phi_0 | H_{\rm c} | \phi_0  \rangle~,
\end{eqnarray}
where $\ket{\phi_0}$ represents the bath state after reset. If the bath is continuously measured and reset, it remains fixed in the state $\ket{\phi_0}$. Consequently, the system experiences an effective mean field from the bath, described by the mean-field Hamiltonian $H_{\rm s0}$. If $\Delta t$ is short, only the bath qubit directly coupled to the system will contribute, as information from other bath qubits does not have enough time to propagate before measurement and reset.

\subsection{Low order Zeno effect}
\label{app:LowZeno}
In the general case, the power is $\alpha=1$. The measurement basis is changed under the operation $H_{{\rm c}0}$, i.e., $\left[H_{{\rm c}0}, |\phi_m \rangle\langle \phi_m| \right]\neq 0$.
We can decompose the system wave function in the orthogonal basis
\begin{eqnarray}
    |\psi_t\rangle=\sum_l a_l|\psi_l\rangle.
\end{eqnarray}

We define the time scale
\begin{eqnarray}
    \omega_{l',m}=\sum_l a_l \langle\phi_{m}| \langle\psi_{l'}|H|\psi_l\rangle|\phi_0\rangle.
\end{eqnarray}
Although $\omega_{l',m}$ depends on the initial system state $|\psi_t\rangle$, we can approximate it by replacing it with its average, i.e., $|\omega_{l',m}|^2 \rightarrow \overline{|\omega_{l',m}|^2}$. In the deep thermalized regime, where $\Delta t$ is large, this average is effectively taken over random states. In contrast, when $\Delta t \rightarrow 0$, measurements occur so frequently that the bath remains effectively frozen in its initial state $|\phi_0\rangle$. In this limit, the system state $|\psi_t\rangle$ can be approximated as evolving under the unitary $e^{-i H_{\rm s 0} t}$, and the average can be performed over pure time evolution.

In the limit of short-time evolution $\Delta t\rightarrow0$, according to the Schrödinger equation, the probability of measuring the bath in the state
$|\phi_m\rangle$ for $m\neq0$ is
\begin{eqnarray}
    P_{m}\approx \Delta t^2\sum_{l'}\overline{|\omega_{l',m}|^2}.
\end{eqnarray}
The probability of the bath remaining in the state $|\phi_0\rangle$ is 
\begin{eqnarray}
    P_0=1-\sum_{m\neq0}P_{m}\approx 1-\Delta t^2\sum_{m\neq0,l'}\overline{|\omega_{l',m}|^2}.
\end{eqnarray}
For a rough lower bound, we retain only one term in
the summation where all measurement results yield $|\phi_0\rangle$. The frame potential then satisfies
\begin{eqnarray}
    F^{(K)}&>&P_0^{2n} \nonumber\\
    &>&{\rm Exp}\left(-\frac{2T^2}{n}\sum_{m\neq0,l'}\overline{|\omega_{l',m}|^2}  \right).
\end{eqnarray}
In this case, the coefficient is $c_H=2\sum_{m\neq0,l'}\overline{|\omega_{l',m}|^2}$.

\subsection{High-order Zeno effect}
\label{app:HighZeno}
In the special case, the power is $\alpha=3$ when the measurement basis ${|\phi_m\rangle}$ is invariant under the first order action of $H_{\rm c0}$, i.e., $\left[H_{\rm c0}, |\phi_m\rangle\langle \phi_m|\right] = 0$. Under this symmetry condition, we obtain a tighter lower bound on the frame potential.

For convenience in the derivation, the bath Hamiltonian is decomposed into a component diagonal in the measurement basis and an off-diagonal term,
\begin{eqnarray}
    H_{{\rm b}}=H_{{\rm bm}}+V ~,
\end{eqnarray}
where $H_{{\rm bm}}$ commutes with the measurement.

In the limit of short time evolution $\Delta t\rightarrow0$,
the unitary could be written as \cite{WangYan2024}
\begin{eqnarray}
    e^{-iH\Delta t}\approx e^{-iH_{{\rm c0}}\frac{\Delta t}{2}}e^{-iH_{{\rm s0}}\Delta t-iH_{{\rm b}}\Delta t}e^{-iH_{{\rm c0}}\frac{\Delta t}{2}}.
\end{eqnarray}
With the initial system state $\ket{\psi_t}$ and bath state $\ket{\phi_0}$, the quantum state after the short time evolution is 
\begin{eqnarray}
|\Psi_{t+\Delta t}\rangle&\approx&	e^{-iH_{{\rm c0}}\frac{\Delta t}{2}}e^{-iH_{{\rm s0}}\Delta t-iH_{{\rm b}}\Delta t}e^{-iH_{{\rm c0}}\frac{\Delta t}{2}}|\psi_{t}\rangle|\phi_{0}\rangle \nonumber\\
&=&	e^{-iH_{{\rm c0}}\frac{\Delta t}{2}}|\psi_{t+\Delta t}\rangle|\phi_{\Delta t}\rangle, 
\end{eqnarray}
where $|\psi\rangle$ represents the system wave function, while $|\phi\rangle $ denotes the bath wave function. 

Expanding the bath wave function in the measurement basis $|\phi_m\rangle$, we have
\begin{eqnarray}
|\Psi_{t+\Delta t}\rangle\approx \sum_m e^{-i\langle \phi_m|H_{\rm c0}| \phi_m\rangle\frac{\Delta t}{2}}c_m|\psi_{t+\Delta t}\rangle|\phi_m\rangle .
\end{eqnarray}

In the first-order time-dependent perturbation theory, we consider only the states directly coupled to the initial state. The corresponding amplitude for $m\neq0$ is given by
\begin{eqnarray}
    c_{m}&\approx&-\frac{1}{E_{0}-E_{m}}\langle \phi_m|V|\phi_0\rangle \left(e^{-i\left(E_{0}-E_{m}\right)\Delta t}-1\right),\nonumber\\
    P_m&\approx&4\left|\frac{\langle \phi_m|V|\phi_0\rangle}{E_{0}-E_{m}}\right|^2 \sin^2\frac{\left(E_{0}-E_{m}\right)\Delta t}{2} ~,
\end{eqnarray}
and $P_0=1-\sum_{m\neq0}P_m$.

The frame potential can be simplified to
\begin{eqnarray}
F^{(k)}&>&\sum_{z,z'}p_z  p_{z'}\left| \langle \psi_z |  \psi_T\rangle \langle \psi_T |\psi_{z'} \rangle \right|^{2K} \nonumber\\
    &>&\left( \sum_{z}p_z\left| \langle \psi_z |  \psi_T\rangle\right|^{2K}  \right)^2,
\label{eq:fp}
\end{eqnarray}
where $|\psi_T \rangle$  is the system wave function when the measurement results remain $|\phi_0 \rangle$ throughout the entire process.

We define the time scale $\mu_m$ as half the difference between the largest and smallest eigenvalues of $\langle\phi_m|H_{\rm c0}|\phi_m\rangle$. The frame potential then simplifies to
\begin{eqnarray}
    F^{(K)} &>& \left(P_0+\sum_{m\neq0} P_m \cos^{2K} \mu_m \Delta t \right)^{2n} \nonumber\\
     &>& \left(1-\sum_m\left|\langle \phi_m|V|\phi_0\rangle\right|^2 K\mu_m^2 \Delta t^4  \right)^{2n} \nonumber\\
    &>&{\rm Exp}\left( -\frac{2KT^4
    }{n^3} \sum_m\left|\langle \phi_m|V|\phi_0\rangle\right|^2 \mu_m^2 \right) ~.
\end{eqnarray}
Because $\mu_m = 0$ whenever $\langle \phi_m|H_{\rm c0}|\phi_m\rangle = 0$, only a small subset of intermediate states contributes. For the Ising Hamiltonian $H_{\rm Ising}$, this expression reduces to $c_H=2J_x^2J_{zz}^2$.

\section{Experiment Methods}
\label{app:Experiment}
\begin{figure*}[tb]
\begin{center}
\includegraphics[clip = true, width =2\columnwidth]{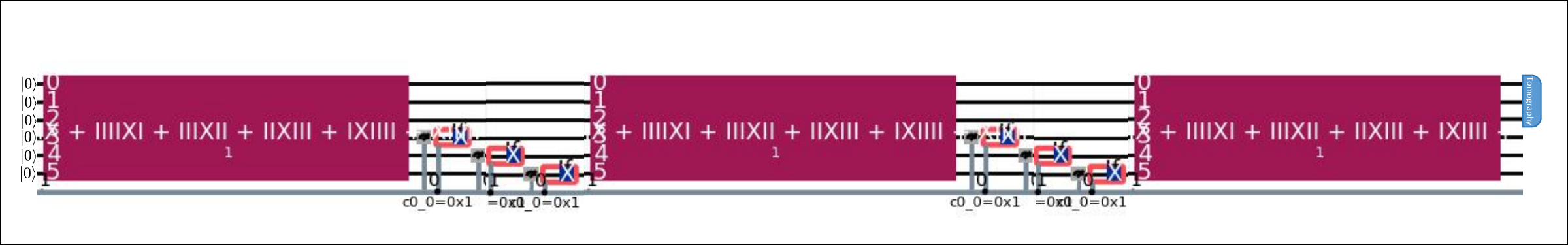}
\caption{\label{fig:ExpCir} Experimental quantum circuit diagram implemented on the IBM Quantum Platform for $n=3$. }
\end{center}
\end{figure*}

We implement our protocol on the IBM Quantum Platform, executing it on the “ibm\_torino’’ QPU. The experimental circuit, shown in Fig.~\ref{fig:ExpCir}, uses $n_{\rm all}=6$ qubits and one classical bit. Among them, $n_{\rm s}=3$ qubits form the system and evolve coherently throughout the process which are tomographed at the end. The remaining $n_{\rm b}=3$ qubits serve as the bath and are measured and reset at every iteration. The classical bit is utilized to enable the reset of the bath after measurements via the following procedure. Each measurement outcome is written to the classical bit: if the outcome is 1, an $X$ gate is applied; otherwise the qubit is left unchanged.

Hamiltonian evolution is implemented using a symmetric Suzuki–Trotter decomposition, which reduces sensitivity to Trotter ordering. In our experiment, the number of Trotter repetitions is 5 for $n=1$, 2 for $n=2$ and 1 for larger $n$ in our experiment.

For small system sizes, we perform full Pauli tomography. The reduced density matrix is expanded as
\begin{eqnarray}
    \rho_{\rm s}=\sum_j p_j\sigma_j ~,
\end{eqnarray}
where each coefficient $p_j$ is obtained by measuring in the corresponding Pauli basis. The first-order frame potential then follows as
\begin{eqnarray}
    F^{(1)}={\rm Tr}(\rho_{\rm s}^2)=N_{\rm s}\sum_j p_j^2 ~.
\end{eqnarray}

For the larger system size, we use classical-shadow tomography. A random local unitary $U_j$ is applied before measuring in the computational basis, yielding a bitstring $|b_j\rangle$. The corresponding snapshot is
\begin{eqnarray}
    \rho_j=(N_{\rm s}+1)U_j^\dagger|b_j\rangle\langle b_j|U_j-I ~,
\end{eqnarray}
which can be computed efficiently. The frame potential is estimated using the unbiased shadow estimator
\begin{eqnarray}
    F^{(1)}={\rm Tr}(\rho_{\rm s}^2)\approx\frac{\sum_{j<k} {\rm Tr}(\rho_j\rho_k)}{N_{\rm samp}(N_{\rm samp}-1)} ~,
\end{eqnarray}
where $N_{\rm samp}$ is the number of snapshots. The snapshot budget is small, satisfying $N_{\rm samp} \ll N_{\rm s}$.

Higher-order frame potentials can, in principle, be accessed by running two identical circuits in parallel and performing a SWAP test to evaluate the overlap between their output states. However, this approach requires an experimental error rate roughly half that of the single-copy protocol.

The noisy simulations presented in Fig.~6 of the manuscript employ the built-in noise model of the IBM Quantum Platform, with all parameters directly extracted from the “ibm\_torino’’ device. A subset of these parameters is listed in Tab.~\ref{tab:table1}. The simulated architecture matches the experimental configuration.

\begin{table}[b]%The best place to locate the table environment is directly after its first reference in text
\caption{\label{tab:table1}%
Selected device parameters of the QPU “ibm\_torino”.
}
\begin{ruledtabular}
\begin{tabular}{ccc}
Parameter& Label &
\multicolumn{1}{c}{\textrm{Median Value}}\\
%\mbox{Three}&\mbox{Four}&\mbox{Five}\\
\hline
Dephasing time &$T_{\rm d}$ & 136 $\mu$s \\
Readout length & $t_{\rm r}$& 1.56 $\mu$s \\
Gate length & $t_{\rm g}$& 68 ns \\
Two-qubit gate error & $\varepsilon_{\rm 2}$& 2.56$\times10^{-3} $ \\
Measure error & $\varepsilon_{\rm m}$&  2.86$\times10^{-3} $ 
\end{tabular}
\end{ruledtabular}
\end{table}

In the noisy intermediate-scale quantum (NISQ) regime, computational resources are severely constrained. Below we present an order-of-magnitude analysis adapted to contemporary superconducting architectures.

The total accumulated error rate $\varepsilon_{\rm a}$ can be obtained from individual errors $\varepsilon_j$, via $1- \varepsilon_{\rm a}=\prod_j \left(1-\varepsilon_j \right)$. Approximately, we have
\begin{eqnarray}
  \ln \left( 1- \varepsilon_{\rm a}\right)&=&\sum_j \ln\left(1-\varepsilon_j \right) \approx-\sum_j \varepsilon_j ~,
\end{eqnarray}
where the index $j$ runs over all relevant noise channels. This linearized form provides a convenient approximation for estimating error budgets in the NISQ regime.

For highly entangled states for $n_{\rm s}$ qubits, the effective coherence time is reduced to
\begin{eqnarray}
    T_{\rm c}\approx \frac{T_{\rm d}}{n_{\rm s}}~.
\end{eqnarray}
The total evolution time required for a Trotterized implementation scales as
\begin{eqnarray}
    T_{\rm Tro}\sim t_{\rm g}n_{\rm all}^\zeta n_{\rm Tro}  ~,
\end{eqnarray}
where $\zeta\in[0,1]$ quantifies the parallelizability of the circuit ($\zeta=0$ for fully sequential execution; $\zeta=1$ for fully parallel execution). The Trotter number is typically constrained by digitalization errors and scales as $n_{\rm Tro}\sim n_{\rm all} J_{zz} T$ \cite{Childs2021,Ostmeyer_2023,Myers2023}, where $n_{\rm all}$ is the total number of qubits.
The total wall-clock duration of the protocol is then
\begin{eqnarray}
    T_{\rm all}&=& T_{\rm Tro}+(n-1)n_{\rm b}^\zeta t_{\rm r}+n_{\rm s}^\zeta t_{\rm r}   ~.
\end{eqnarray}

Gate errors accumulate approximately as
\begin{eqnarray}
    \varepsilon_{\rm g} \sim \varepsilon_2 n_{\rm all}n_{\rm Tro} ~,
\end{eqnarray}
where single-qubit gate errors are neglected.

The measurement and reset error of the bath scales as $(n-1)n_{\rm b}$, while the readout error of the system is $n_{\rm s}\varepsilon_{\rm m}$.

Combining all contributions, the total error budget satisfies
\begin{eqnarray}
    \ln \left( 1- \varepsilon_{\rm a}\right) &\approx&  -\frac{ T_{\rm all}}{ T_{\rm c}}-\varepsilon_{\rm g}-\left((n-1)n_{\rm b} +n_{\rm s} \right)\varepsilon_{\rm m} .
\end{eqnarray}
The first term originates from the finite coherence time, the second from accumulated gate errors, and the third from measurement errors. In the highly parallel operation $\zeta=0$ and small number of mid-circuit measurement $n\rightarrow1$, the total error is dominated by the accumulated gate errors. 

For first-order Suzuki–Trotter evolution under $H_{\rm Ising}$, the accumulated error satisfies
\begin{equation}
\varepsilon_{\rm a}\gtrsim 2\varepsilon_{2} n_{\rm all}^2 J_{zz}T
\end{equation} 
or equivalently
\begin{equation}
    J_{zz}T \lesssim\frac{\varepsilon_{\rm a}}{ 2n_{\rm all}^2\varepsilon_{2}}.
\end{equation}
Substituting the experimental parameters yields the constraint $J_{zz}T \lesssim 5$. Combining this with Eq. (12) of the manuscript then implies that the optimal number of mid-circuit measurements is bounded by $n_\gamma \lesssim 4$.

Future improvements in random-state generation will require reduced gate error rates, shorter gate durations, and longer qubit coherence times. Most importantly, native Hamiltonian evolution and parallelized gate execution at the hardware level would significantly loosen these constraints.

\bibliography{myref}

\end{document}